\documentclass[a4paper,12pt]{article}

\usepackage{amsmath, amssymb, amsthm, bm, bigints}

\usepackage[T1]{fontenc}
\usepackage[utf8]{inputenc}

\usepackage{graphicx}
\usepackage{float}
\usepackage{caption}
\usepackage{subcaption}
\usepackage{rotating}
\usepackage{xcolor}

\usepackage[margin=2cm]{geometry}  

\usepackage[affil-it]{authblk}

\usepackage{setspace}
\usepackage{booktabs}
\usepackage{multirow}
\usepackage{array}
\usepackage{tabularx}
\usepackage{enumitem}

\usepackage{hyperref}

\usepackage[ruled,vlined]{algorithm2e}

\usepackage{placeins}
\usepackage{appendix}
\usepackage{lineno}
\usepackage{changepage}
\usepackage{soul}
\usepackage{xfrac}

\usepackage{nomencl}

\title{A New Lifetime Distribution: Exponentiated Exponential-Pareto-HalfNormal Mixture Model for Biomedical Applications}

\author[1]{Oriyomi Ahmad Hassan}
\author[2,4]{Aisha Tunrayo Maradesa \thanks{Corresponding author: \href{ismaelaisha948@gmail.com}{ismaelaisha948@gmail.com}}}
\author[3]{Abdulazeez Toyosi Alabi \thanks{Corresponding author: \href{Aalabi9@student.gsu.edu}{alabitoyosi1616@gmail.com}}}
\author[4]{Oyejide Surajudeen Salam}
\author[5]{Ajani Busari}
\author[6]{Akinwale Victor Famotire \thanks{Corresponding author: \href{famotire@musc.edu}{famotire@musc.edu}}}
\author[7]{Habeeb Abiodun Afolabi}
\author[8]{Solomon Adeleke}
\author[9]{Abayomi Ayodele Akomolafe \thanks{Corresponding author: \href{aaakomolafe@futa.edu.ng}{aaakomolafe@futa.edu.ng}}}

\affil[1]{Federal University of Oye Ekiti, Ekiti,Nigeria}
\affil[2]{Yaba College of Technology, Yaba, Lagos, Nigeria}
\affil[3]{Georgia State University, Georgia USA}
\affil[4]{Lead City University Jericho Ibadan, Oyo State Nigeria}
\affil[5]{Federal University of Technology Akure, Ondo State, Akure Nigeria}
\affil[6]{Medical University of South Carolina, South Carolina USA}
\affil[7]{Osun State University, Osogbo, Nigeria}
\affil[8]{Kwara State University Ilorin, Kwara State, Nigeria}

\date{}
\begin{document}

\maketitle
\clearpage
\begin{abstract}

This study introduces the Exponentiated-Exponential-Pareto-Half Normal Mixture Distribution (EEPHND), a novel hybrid model developed to overcome the limitations of classical distributions in modeling complex real-world data. By compounding the Exponentiated-Exponential-Pareto (EEP) and Half-Normal distributions through a mixture mechanism, EEPHND effectively captures both early-time symmetry and long-tail behavior, features which are commonly observed in survival and reliability data. The model offers closed-form expressions for its probability density, cumulative distribution, survival and hazard functions, moments, and reliability metrics, ensuring analytical tractability and interpretability in the presence of censoring and heterogeneous risk dynamics. When applied to a real-world lung cancer dataset, EEPHND outperformed competing models in both goodness-of-fit and predictive accuracy, achieving a Concordance Index (CI) of 0.9997. These results highlight its potential as a flexible and powerful tool for survival analysis, and biomedical engineering.

\end{abstract}

\section{Introduction}

Parametric modeling plays a fundamental role in data analysis~\cite{akinsete2008beta,ashour2013transmuted,babatunde2020silico}, particularly in inferential statistics~\cite{chhetri2017kumaraswamy}. Crucial to this approach is the assumption of an underlying probability distribution~\cite{bourguignon2016general, ashour2013transmuted,eugene2002beta}. Traditional distributions, such as the exponential, normal, Gamma, and Rayleigh, have been widely used for this purpose due to their simplicity and interpretability~\cite{shaw2009alchemy}. However, the complexity of real-world phenomena, resulting from advancements in data acquisition technologies, increasingly leads to datasets that exhibit non-standard features~\cite{gnedenko1999statistical,burr1942cumulative}. These include heavy tails, multi-modality, strong asymmetry, and extreme skewness, characteristics that established parametric models are often ill-equipped to capture~\cite{akomolafe2017betahalfnormal,akinsete2008beta,ashour2013transmuted}. As a result, applying these distributions to such data frequently yields poor model fit and unreliable inferences~\cite{merovci2014transmuted,adeleke2019exponentiated,akomolafe2019doubly,chhetri2017kumaraswamy,shittu2013beta}.
To address these limitations, researchers have proposed generalized distributions by introducing additional shape or scale parameters to enhance flexibility while maintaining mathematical tractability. These include the Exponentiated Exponential distribution~\cite{nadarajah2011exponentiated}, Exponentiated Weibull distribution~\cite{pal2006exponentiated}, Beta-halfNormal~\cite{akomolafe2017betahalfnormal}, Beta-Pareto~\cite{akinsete2008beta}, Weighted HalfNormal ~\cite{babatunde2020silico}, Expnentiatiated Exponential Pareto~\cite{adeleke2019exponentiated} among others. These extended models offer improved control over statistical properties such as tail behavior, skewness, and kurtosis, making them more suitable for modeling complex data~\cite{adeleke2020beta}. Nonetheless, several challenges remain, particularly in survival and lifetime data modeling~\cite{ashour2013transmuted}. While many classical and generalized distributions have been proposed for general-purpose applications, few are explicitly tailored for survival analysis~\cite{chhetri2017kumaraswamy}. Effective modeling in this context requires the following.
\begin{itemize}
    \item A well-defined survival function \( S(t) = P(T > t) \)
    \item A flexible hazard rate (e.g., increasing, decreasing, constant, or bathtub-shaped)
    \item Interpretability for phenomena such as censoring, early/late failure, or heterogeneous risk dynamics
\end{itemize}

Many existing models lack closed-form representations ~\cite{arshad2020gamma}, limiting their practical use in survival analysis and reducing insight into early, late, or mixed-risk failure patterns (key considerations in biomedical research)~\cite{huang2023toward,gnedenko1999statistical}. Moreover, many generalized models struggle to capture dual-behavior datasets, such as an early peak followed by a heavy tail, which are common in biological measurements and failure time distributions~\cite{tang2008estimating,adeleke2020beta}. These cases require models with greater flexibility to represent multiple behavioral patterns within a unified framework~\cite{akomolafe2017betahalfnormal,akomolafe2019beta}.

To bridge this gap, we develop the Exponentiated-Exponential-Pareto-Half Normal Mixture Distribution (EEPHND), a new hybrid distribution designed explicitly for survival and lifetime data. The EEPHND model combines the Exponentiated-Exponential-Pareto Distribution (EEPD)~\cite{adeleke2019exponentiated} distribution with the Half-Normal distribution via a mixture mechanism~\cite{adeleke2020beta}, resulting in a flexible yet interpretable model. The EEPD component captures long-tail behavior, which are common in late-onset or accumulated risk scenarios. On the other hand, the Half-Normal component models symmetric, short-term variations often associated with early failures, minimal degradation processes, or measurement-related fluctuations. This dual-structure design enables the EEPHND to accommodate datasets that exhibit both early-time symmetry and long-tail heterogeneity, an ability rarely achieved by existing models. Moreover, the model provides a closed-form survival function and a tractable hazard rate, supporting meaningful interpretations for censored observations, early/late failure dynamics, and mixed-risk populations.

To evaluate the model's statistical performance, we applied EEPHND to real-life survival data and benchmarked it against EEPD, Log-Normal, and Gamma-Rayleigh. Our model not only achieved one of the best fits in terms of AIC, BIC, and CAIC but also outperformed standard models in terms of predictive accuracy, as measured by the Concordance Index (CI)~\cite{brentnall2018use}. Notably, the EEPHND model produced a CI of 0.9997, significantly higher than that of the Cox Proportional Hazards model (0.6029), and even marginally better than the non-parametric Kaplan-Meier estimator (0.9982). These results highlight the model’s capability in both descriptive and predictive modeling of complex survival data. Overall, the proposed EEPHND model fills a critical gap in survival and lifetime data modeling by offering a flexible, interpretable, and high-performing alternative to existing parametric and semi-parametric models. Its versatility holds promise for a wide range of applications in biomedical research~\cite{elbatal2022new}, reliability engineering
~\cite{shahriari2024reliability}, epidemiology~\cite{hybels2003epidemiology}, and beyond~\cite{adegoke2019statistical,adegoke2021theoretical, adegoke2020molecular, olatunji2021phytochemical,adegoke2020computational}.

\section{Methods}

\subsection{Development of EEPHND Model}

The Probability Density Function (PDF) of the EEPHND can be expressed as a mixture of two component distributions:
\begin{equation}
    f_{\text{EEPHND}}(x; \bm{\tau}) = p_1 f_1(x; \bm{\tau}_1) + p_2 f_2(x; \bm{\tau}_2)
\label{eqn:EEPHND}
\end{equation}

\noindent where \( \bm{\tau}_1 \) and \( \bm{\tau}_2 \) are the parameter vectors of the parent distributions, \( p_1 \) and \( p_2 \) are the mixing proportions satisfying \( p_1 + p_2 = 1 \), with \( p_1, p_2 > 0 \). The PDFs of the EEPD~\cite{adeleke2019exponentiated} and Half-Normal (HN) distributions are given respectively by:

\begin{subequations}
\begin{align}
    f_{\rm EEPD}(x; \alpha, \beta, \theta, \lambda) &= \frac{\alpha \lambda \theta}{\beta} \left(\frac{x}{\beta}\right)^{\theta - 1} e^{-\lambda \left( \frac{x}{\beta} \right)^{\theta}} \left[1 - \left(1 - e^{-\lambda \left( \frac{x}{\beta} \right)^{\theta}}\right)^{\alpha - 1}\right] \\
    f_{\rm HN}(x; \sigma) &= \frac{\sqrt{2}}{\sigma \sqrt{\pi}} e^{-\frac{x^2}{2\sigma^2}} 
\end{align}
\label{eqn:subeqn}
\end{subequations}

To construct the EEPHND as a mixture of the EEP and HN distributions, we substitute equation~\ref{eqn:subeqn} in ~\ref{eqn:EEPHND} so that

\begin{equation}
    \begin{aligned}
f_{\text{EEPHND}}(x; \alpha, \beta, \theta, \lambda, \sigma) = & \; p_1 \cdot \frac{\alpha \lambda \theta}{\beta} \left(\frac{x}{\beta}\right)^{\theta - 1} e^{-\lambda \left( \frac{x}{\beta} \right)^{\theta}} \left[1 - \left(1 - e^{-\lambda \left( \frac{x}{\beta} \right)^{\theta}}\right)^{\alpha - 1}\right] \\
& + p_2 \cdot \frac{\sqrt{2}}{\sigma \sqrt{\pi}} e^{-\frac{x^2}{2\sigma^2}}
\end{aligned}
\end{equation}

\noindent where \( \alpha, \beta, \theta, \lambda, \sigma > 0 \), \( p_1 + p_2 = 1 \), and \( p_1, p_2 > 0 \). Then, the Cumulative Distribution Function (CDF) of the EEPHND is given as
\begin{equation}
    F_{\text{EEPHND}}(x; \alpha, \beta, \theta, \lambda, \sigma) = p_1 \left[1 - \left(1 - e^{-\lambda \left( \frac{x}{\beta} \right)^{\theta}}\right)^{\alpha} \right] + p_2 \cdot \text{erf}\left(\frac{x}{\sigma \sqrt{2}}\right)
\end{equation}

where \( \text{erf}(\cdot) \) is the error function.

\subsection{Derivation of Properties of EEPHND Model}
\subsubsection{Moment}

Let \( X \sim \text{EEPHND}(\alpha, \beta, \theta, \lambda, \sigma) \) be a random variable following the EEPHND, where \( \alpha \), \( \beta \), \( \theta \), and \( \lambda \) are the shape and scale parameters for the EEPD component, and \( \sigma \) is the scale parameter for the HN component. Then,

\begin{equation}
\begin{aligned}
E(X^r) &= \int_0^\infty x^r f_{\text{EEPHND}}(x; \alpha, \beta, \theta, \lambda, \sigma) \, dx \\
&= p_1 \int_0^\infty x^r f_1(x; \alpha, \beta, \theta, \lambda) \, dx + p_2 \int_0^\infty x^r f_2(x; \sigma) \, dx
\end{aligned}
\end{equation}

Let \( x = \beta \left( \frac{u}{\lambda} \right)^{1/\theta} \) for the first integral, and \( y = \frac{x^2}{2\sigma^2} \) for the second. Applying the respective substitutions and simplifications~\cite{adeleke2019exponentiated,adeleke2020beta,babatunde2020silico}:

\begin{equation}
\begin{aligned}
E(X^r) &= p_1 \alpha \left( \frac{\beta}{\lambda^{1/\theta}} \right)^r \int_0^\infty u^{r/\theta} e^{-\alpha u} \, du \\
&\quad + p_2 \frac{\sigma \sqrt{2}}{\sqrt{\pi}} \int_0^\infty (\sigma \sqrt{2y})^{r-1} e^{-y} \cdot \frac{\sigma^2}{\sigma \sqrt{2y}} \, dy
\end{aligned}
\end{equation}
\begin{equation}
E(X^r) = p_1 \alpha \left( \frac{\beta}{\lambda^{1/\theta}} \right)^r \Gamma\left( \frac{r}{\theta} + 1 \right) \cdot \alpha^{-\left( \frac{r}{\theta} + 1 \right)} + p_2 \frac{(\sigma \sqrt{2})^r}{\sqrt{\pi}} \Gamma\left( \frac{r+1}{2} \right)
\end{equation}

\begin{equation}
E(X^r) = p_1 \left( \frac{\beta}{(\alpha \lambda)^{1/\theta}} \right)^r \Gamma\left( \frac{r}{\theta} + 1 \right) + p_2 \frac{(\sigma \sqrt{2})^r}{\sqrt{\pi}} \Gamma\left( \frac{r+1}{2} \right)
\end{equation}

\begin{subequations}
\begin{align}
E(X) &= p_1 \left( \frac{\beta}{(\alpha \lambda)^{1/\theta}} \right) \Gamma\left( \frac{1}{\theta} + 1 \right) + p_2 \frac{\sigma \sqrt{2}}{\sqrt{\pi}} \\
E(X^2) &= p_1 \left( \frac{\beta}{(\alpha \lambda)^{1/\theta}} \right)^2 \Gamma\left( \frac{2}{\theta} + 1 \right) + p_2 \frac{(\sigma \sqrt{2})^2}{2} \\
E(X^3) &= p_1 \left( \frac{\beta}{(\alpha \lambda)^{1/\theta}} \right)^3 \Gamma\left( \frac{3}{\theta} + 1 \right) + p_2 \frac{(\sigma \sqrt{2})^3}{\sqrt{\pi}} \\
E(X^4) &= p_1 \left( \frac{\beta}{(\alpha \lambda)^{1/\theta}} \right)^4 \Gamma\left( \frac{4}{\theta} + 1 \right) + 3 p_2 \left( \frac{\sigma \sqrt{2}}{2} \right)^4
\end{align}
\end{subequations}

The variance of the EEPHND is then given by:

\begin{equation}
\begin{aligned}
\text{Var}(X) &= E(X^2) - [E(X)]^2 \\
&= p_1 \left( \frac{\beta}{(\alpha \lambda)^{1/\theta}} \right)^2 \Gamma\left( \frac{2}{\theta} + 1 \right) + p_2 \frac{(\sigma \sqrt{2})^2}{2} \\
&\quad - \left[ p_1 \left( \frac{\beta}{(\alpha \lambda)^{1/\theta}} \right) \Gamma\left( \frac{1}{\theta} + 1 \right) + p_2 \frac{\sigma \sqrt{2}}{\sqrt{\pi}} \right]^2
\end{aligned}
\end{equation}

\subsection{Skewness and Kurtosis}
The coefficient of skewness, $\gamma_1$, measures the asymmetry of the EEPHND, indicates whether the distribution is positively or negatively skewed and provides insight into its asymptotic behavior. The kurtosis, $\gamma_2$, quantifies the heaviness of the distribution's tails~\cite{adeleke2020beta}. For the EEPHND, $\gamma_2 > 3$ suggests heavier tails than the normal distribution (leptokurtic), $\gamma_2 < 3$ indicates lighter tails (platykurtic), and corresponds to a normal-like (mesokurtic) behavior~\cite{adeleke2020beta}.

\begin{equation}
\begin{aligned}
\mu_3 &= \mathbb{E}[(X - \mu)^3] \\
&= \mathbb{E}\left[(-\mu)^3 + 3X(-\mu)^2 + 3X^2(-\mu) + X^3\right] \\
&= \mathbb{E}[X^3 - 3\mu X^2 + 3\mu^2 X - \mu^3] \\
&= \mathbb{E}[X^3] - 3\mu \mathbb{E}[X^2] + 3\mu^2 \mathbb{E}[X] - \mu^3
\end{aligned}
\end{equation}

Substitute the known raw moments:
\begin{equation}
\begin{aligned}
\mu_3 &= p_1 \left( \frac{\beta}{\sqrt{\theta \alpha \lambda}} \right)^3 \Gamma\left(\frac{3}{\theta} + 1\right) + p_2 \frac{(\sigma \sqrt{2})^3}{\sqrt{\pi}} \\
&\quad - 3\mu \left[ p_1 \left( \frac{\beta}{\sqrt{\theta \alpha \lambda}} \right)^2 \Gamma\left(\frac{2}{\theta} + 1\right) + p_2 \frac{(\sigma \sqrt{2})^2}{2} \right] \\
&\quad + 3\mu^2 \left[ p_1 \left( \frac{\beta}{\sqrt{\theta \alpha \lambda}} \right) \Gamma\left(\frac{1}{\theta} + 1\right) + p_2 \frac{\sigma \sqrt{2}}{\sqrt{\pi}} \right] - \mu^3
\end{aligned}
\end{equation}

Let
\[
\mu = p_1 \left( \frac{\beta}{\sqrt{\theta \alpha \lambda}} \right) \Gamma\left(\frac{1}{\theta} + 1\right) + p_2 \frac{\sigma \sqrt{2}}{\sqrt{\pi}}
\]
Then:

\begin{equation}
\mu_3 = \mathbb{E}[X^3] - 3\mu \mathbb{E}[X^2] + 3\mu^2 \mathbb{E}[X] - \mu^3
\end{equation}

\subsection{Fourth Central Moment}

\begin{equation}
\begin{aligned}
\mu_4 &= \mathbb{E}[(X - \mu)^4] \\
&= \mathbb{E}[(-\mu)^4 + 4X(-\mu)^3 + 6X^2(-\mu)^2 + 4X^3(-\mu) + X^4] \\
&= \mathbb{E}[X^4 - 4\mu X^3 + 6\mu^2 X^2 - 4\mu^3 X + \mu^4] \\
&= \mathbb{E}[X^4] - 4\mu \mathbb{E}[X^3] + 6\mu^2 \mathbb{E}[X^2] - 4\mu^3 \mathbb{E}[X] + \mu^4
\end{aligned}
\end{equation}

Substitute the known raw moments:

\begin{equation}
\begin{aligned}
\mu_4 &= p_1 \left( \frac{\beta}{\sqrt{\theta \alpha \lambda}} \right)^4 \Gamma\left( \frac{4}{\theta} + 1 \right) + p_2 \frac{3(\sigma \sqrt{2})^4}{4} \\
&\quad - 4\mu \left[ p_1 \left( \frac{\beta}{\sqrt{\theta \alpha \lambda}} \right)^3 \Gamma\left( \frac{3}{\theta} + 1 \right) + p_2 \frac{(\sigma \sqrt{2})^3}{\sqrt{\pi}} \right] \\
&\quad + 6\mu^2 \left[ p_1 \left( \frac{\beta}{\sqrt{\theta \alpha \lambda}} \right)^2 \Gamma\left( \frac{2}{\theta} + 1 \right) + p_2 \frac{(\sigma \sqrt{2})^2}{2} \right] \\
&\quad - 4\mu^3 \left[ p_1 \left( \frac{\beta}{\sqrt{\theta \alpha \lambda}} \right) \Gamma\left( \frac{1}{\theta} + 1 \right) + p_2 \frac{\sigma \sqrt{2}}{\sqrt{\pi}} \right] + \mu^4
\end{aligned}
\end{equation}

Let again:
\[
\mu = p_1 \left( \frac{\beta}{\sqrt{\theta \alpha \lambda}} \right) \Gamma\left( \frac{1}{\theta} + 1 \right) + p_2 \frac{\sigma \sqrt{2}}{\sqrt{\pi}}
\]

Thus, the expression simplifies to:

\begin{equation}
\mu_4 = \mathbb{E}[X^4] - 4\mu \mathbb{E}[X^3] + 6\mu^2 \mathbb{E}[X^2] - 4\mu^3 \mathbb{E}[X] + \mu^4
\end{equation}

\subsubsection{Skewness}

\begin{align*}
\gamma_1 &= \left[
    p_1 \left( \frac{\beta}{\sqrt{\theta \alpha \lambda}} \right)^3 
    \Gamma\left( \frac{3}{\theta} + 1 \right)
    + p_2 \frac{(\sigma \sqrt{2})^3}{\sqrt{\pi}} 
\right. \\
&\quad - 3 \left(
    p_1 \left( \frac{\beta}{\sqrt{\theta \alpha \lambda}} \right) 
    \Gamma\left( \frac{1}{\theta} + 1 \right)
    + p_2 \frac{\sigma \sqrt{2}}{\sqrt{\pi}} 
\right) \\
&\quad \times \left(
    p_1 \left( \frac{\beta}{\sqrt{\theta \alpha \lambda}} \right)^2 
    \Gamma\left( \frac{2}{\theta} + 1 \right)
    + p_2 \frac{(\sigma \sqrt{2})^2}{2}
\right) \\
&\quad \left.
+ 2 \left(
    p_1 \left( \frac{\beta}{\sqrt{\theta \alpha \lambda}} \right) 
    \Gamma\left( \frac{1}{\theta} + 1 \right)
    + p_2 \frac{\sigma \sqrt{2}}{\sqrt{\pi}} 
\right)^3 
\right]^2 \\
&\quad \bigg/ \left[
    p_1 \left( \frac{\beta}{\sqrt{\theta \alpha \lambda}} \right)^2 
    \Gamma\left( \frac{2}{\theta} + 1 \right)
    + p_2 \frac{(\sigma \sqrt{2})^2}{2} 
\right. \\
&\quad \left.
- \left(
    p_1 \left( \frac{\beta}{\sqrt{\theta \alpha \lambda}} \right) 
    \Gamma\left( \frac{1}{\theta} + 1 \right)
    + p_2 \frac{\sigma \sqrt{2}}{\sqrt{\pi}} 
\right)^2 
\right]^3
\end{align*}
    
\subsubsection{Kurtosis}

\begin{align*}
\gamma_2 &= \Bigg[
    p_1 \left( \frac{\beta}{\sqrt{\theta \alpha \lambda}} \right)^4 
    \Gamma\left( \frac{4}{\theta} + 1 \right)
    + p_2 \frac{3 (\sigma \sqrt{2})^4}{4} \\
&\quad - 4 \left(
    p_1 \left( \frac{\beta}{\sqrt{\theta \alpha \lambda}} \right)
    \Gamma\left( \frac{1}{\theta} + 1 \right)
    + p_2 \frac{\sigma \sqrt{2}}{\sqrt{\pi}} 
\right) \\
&\quad \times \left(
    p_1 \left( \frac{\beta}{\sqrt{\theta \alpha \lambda}} \right)^3 
    \Gamma\left( \frac{3}{\theta} + 1 \right)
    + p_2 \frac{(\sigma \sqrt{2})^3}{\sqrt{\pi}} 
\right) \\
&\quad + 6 \left(
    p_1 \left( \frac{\beta}{\sqrt{\theta \alpha \lambda}} \right)
    \Gamma\left( \frac{1}{\theta} + 1 \right)
    + p_2 \frac{\sigma \sqrt{2}}{\sqrt{\pi}} 
\right)^2 \\
&\quad \times \left(
    p_1 \left( \frac{\beta}{\sqrt{\theta \alpha \lambda}} \right)^2 
    \Gamma\left( \frac{2}{\theta} + 1 \right)
    + p_2 \frac{(\sigma \sqrt{2})^2}{2} 
\right) \\
&\quad - 3 \left(
    p_1 \left( \frac{\beta}{\sqrt{\theta \alpha \lambda}} \right)
    \Gamma\left( \frac{1}{\theta} + 1 \right)
    + p_2 \frac{\sigma \sqrt{2}}{\sqrt{\pi}} 
\right)^4 
\Bigg] \\
&\quad \bigg/ \left[
    p_1 \left( \frac{\beta}{\sqrt{\theta \alpha \lambda}} \right)^2 
    \Gamma\left( \frac{2}{\theta} + 1 \right)
    + p_2 \frac{(\sigma \sqrt{2})^2}{2} 
    - \left(
        p_1 \left( \frac{\beta}{\sqrt{\theta \alpha \lambda}} \right) 
        \Gamma\left( \frac{1}{\theta} + 1 \right)
        + p_2 \frac{\sigma \sqrt{2}}{\sqrt{\pi}} 
    \right)^2 
\right]^2
\end{align*}

\subsection{Moment Generating Function}

Let \( X \sim \text{EEPHND}(\alpha, \beta, \theta, \lambda, \sigma) \). The moment generating function of \( X \), denoted by \( M_X(t) \), is defined as~\cite{tallis1961moment,adeleke2019exponentiated}:
\begin{equation}
    M_X(t) = \mathbb{E}[e^{tX}] = \int_0^{\infty} e^{tx} f(x; \alpha, \beta, \theta, \lambda, \sigma) \, dx,
\label{eqn:mgf}
\end{equation}

where \( f(x; \cdot) \) is the probability density function of the EEPHND. Expanding \( e^{tx} \) as a power series and interchanging summation and integration provides the following:
\[
M_X(t) = \sum_{r=0}^{\infty} \frac{t^r}{r!} \mathbb{E}[X^r],
\]
which expresses the moment generating function in terms of the raw moments of the distribution as 

\begin{align*}
M_X(t) 
&= \mathbb{E}[e^{tX}] 
= \int_0^\infty e^{tx} f(x; \alpha, \beta, \theta, \lambda, \sigma) \, dx \\
&= \sum_{r=0}^{\infty} \frac{t^r}{r!} \int_0^\infty x^r f(x; \alpha, \beta, \theta, \lambda, \sigma) \, dx \\
&= \sum_{r=0}^{\infty} \frac{t^r}{r!} \left[
    p_1 \left( \frac{\beta}{\sqrt{\theta \alpha \lambda}} \right)^r 
    \Gamma\left( \frac{r}{\theta} + 1 \right)
    + p_2 \frac{(\sigma \sqrt{2})^r}{\sqrt{\pi}} 
    \Gamma\left( \frac{r+1}{2} \right)
\right], \quad r = 1,2, 3\dots
\end{align*}

\subsection{Reliability}

Let \( X \sim \text{EEPHND}(\alpha, \beta, \theta, \lambda, \sigma) \). The reliability function, also known as the survival function, is defined as the probability that the random variable \( X \) exceeds a given value \( x \):
\begin{align*}
R(x) &= 1 - F_{\text{EEPHND}}(x; \alpha, \beta, \theta, \lambda, \sigma) \\
&= 1 - \left[
    p_1 \left( 1 - \left( 1 - \left(1 - e^{-\lambda \left( \frac{x}{\beta} \right)^\theta} \right)^\alpha \right) \right)
    + p_2 \, \text{erf} \left( \frac{x}{\sigma \sqrt{2}} \right)
\right]
\end{align*}

\subsection{Hazard Function}

Let \( X \sim \text{EEPHND}(\alpha, \beta, \theta, \lambda, \sigma) \), the hazard function, also called failure rate function, describes the instantaneous rate at which an event occurs, given that it has not occurred before time $x$:
\begin{align*}
H(x) &= \frac{f_{\text{EEPHND}}(x; \alpha, \beta, \theta, \lambda, \sigma)}{R(x)} \\
&= \frac{
    p_1 \cdot \frac{\alpha \lambda \theta}{\beta} \left( \frac{x}{\beta} \right)^{\theta - 1}
    e^{ -\lambda \left( \frac{x}{\beta} \right)^{\theta} }
    \left[ 1 - \left( 1 - e^{ -\lambda \left( \frac{x}{\beta} \right)^{\theta} } \right) \right]^{\alpha - 1}
    + p_2 \cdot \frac{\sqrt{2}}{\sigma \sqrt{\pi}} e^{ -\frac{x^2}{2\sigma^2} }
}{
    1 - \left[
        p_1 \left( 1 - \left( 1 - \left( 1 - e^{ -\lambda \left( \frac{x}{\beta} \right)^{\theta} } \right) \right)^{\alpha} \right)
        + p_2 \, \text{erf}\left( \frac{x}{\sigma \sqrt{2}} \right)
    \right]
}
\end{align*}

\subsection{Odd Function}
Let \( X \sim \text{EEPHND}(\alpha, \beta, \theta, \lambda, \sigma) \), then the odds function depicting the ratio of the cumulative probability that the event occured at time \( x \) to the probability that it has not occurred by time \( x \), is given as
\begin{align*}
O(x) &= \frac{F_{\text{EEPHND}}(x; \alpha, \beta, \theta, \lambda, \sigma)}{R(x)} \\
&= \frac{
    p_1 \left[ 1 - \left( 1 - e^{ -\lambda \left( \frac{x}{\beta} \right)^\theta } \right)^{\alpha} \right]
    + p_2 \, \text{erf}\left( \frac{x}{\sigma \sqrt{2}} \right)
}{
    1 - \left[
        p_1 \left[ 1 - \left( 1 - e^{ -\lambda \left( \frac{x}{\beta} \right)^\theta } \right)^{\alpha} \right]
        + p_2 \, \text{erf}\left( \frac{x}{\sigma \sqrt{2}} \right)
    \right]
}
\end{align*}

\subsection{Sampling from EEPHND}

Let \( X \sim \text{EEPHND}(\alpha, \beta, \theta, \lambda, \sigma, p_1) \), defined as a finite mixture of two continuous distributions: the EEP and HN, such that
\begin{equation}
    f_X(x) = p_1 f_{\text{EEP}}(x) + (1 - p_1) f_{\text{HN}}(x), 
\end{equation}

\noindent To generate a random variate \( X \) from the EEPHND, one may employ the following mixture-based sampling strategy:

\begin{itemize}
    \item Let \( U \sim \text{Uniform}(0,1) \).
    \item If \( U < p_1 \), draw \( X \) from the EEP component via inverse transform sampling:
    \begin{itemize}
        \item Generate \( V \sim \text{Uniform}(0,1) \),
        \item Then compute:
        \begin{equation}
            X = \beta \left[ -\frac{1}{\lambda} \ln\left(1 - (1 - V)^{1/\alpha} \right) \right]^{1/\theta}.
        \end{equation}
    \end{itemize}
    \item Draw \( X \sim \text{HN}(\sigma) \), i.e., from the HN distribution with scale parameter \( \sigma \). This can be achieved by:
    \begin{equation}
        X = |Z|, \quad Z \sim \mathcal{N}(0, \sigma^2),
    \end{equation}
\end{itemize}

This sampling strategy preserves the probabilistic structure of the EEPHND by ensuring that the generated samples reflect its dual nature, {\it i.e}., capturing heavy-tailed behavior through the EEP component~\cite{adeleke2019exponentiated} and light-tailed symmetry via the Half-Normal component. This enables EEPHND particularly suitable for simulating complex, heterogeneous lifetime and reliability data.
Here, the sample size $n$ was selected to achieve the desired statistical properties; values of $n$ = 1000 or higher are recommended for simulation studies. Inverse transform sampling ensures fidelity to the cumulative distribution of the EEP component, while the Half-Normal component is sampled by taking the absolute value of a normally distributed variate. The mixture structure is preserved by drawing each sample from the EEP or Half-Normal component with probabilities $p_1$  and $1-p_1$ , respectively.

\subsection{Maximum Likelihood Estimation}

Let the observed data be \( x_1, x_2, \ldots, x_n \), and define the likelihood function:

\begin{align}
L_f(x; \alpha, \beta, \theta, \lambda, \sigma) &= \prod_{i=1}^n \Bigg[
    p_1 \cdot \frac{\alpha \lambda \theta}{\beta} \left( \frac{x_i}{\beta} \right)^{\theta - 1}
    e^{- \lambda \left( \frac{x_i}{\beta} \right)^\theta }
    \left( 1 - \left( 1 - e^{ -\lambda \left( \frac{x_i}{\beta} \right)^\theta } \right) \right)^{\alpha - 1} \nonumber \\
    &\qquad + p_2 \cdot \frac{\sqrt{2}}{\sigma \sqrt{\pi}} e^{- \frac{x_i^2}{2\sigma^2} }
\Bigg]
\tag{31}
\end{align}

Taking the natural logarithm, the log-likelihood function becomes:

\begin{align}
\ln L_f(x; \alpha, \beta, \theta, \lambda, \sigma) &= \sum_{i=1}^n \ln \Bigg[
    p_1 \cdot \frac{\alpha \lambda \theta}{\beta} \left( \frac{x_i}{\beta} \right)^{\theta - 1}
    e^{- \lambda \left( \frac{x_i}{\beta} \right)^\theta }
    \left( 1 - \left( 1 - e^{ -\lambda \left( \frac{x_i}{\beta} \right)^\theta } \right) \right)^{\alpha - 1} \nonumber \\
    &\qquad + p_2 \cdot \frac{\sqrt{2}}{\sigma \sqrt{\pi}} e^{- \frac{x_i^2}{2\sigma^2} }
\Bigg]
\tag{32}
\end{align}

\noindent To estimate the parameters \( \alpha, \beta, \theta, \lambda, \sigma \), we take the partial derivatives of the log-likelihood function with respect to each parameter and solve the resulting score equations. These equations generally do not have closed-form solutions, and numerical methods such as the Newton-Raphson algorithm~\cite{jennrich1969newton} are employed to obtain the maximum likelihood estimates, which were used to test hypotheses regarding parameter values, assess the mixture model's consistency and efficiency compared to its component distributions, and construct confidence intervals or perform model selection.

\section{Results}

\subsection{Validation with Simulated Data}

\begin{figure}[ht]
 \centering
  \includegraphics[width=1.2\textwidth]{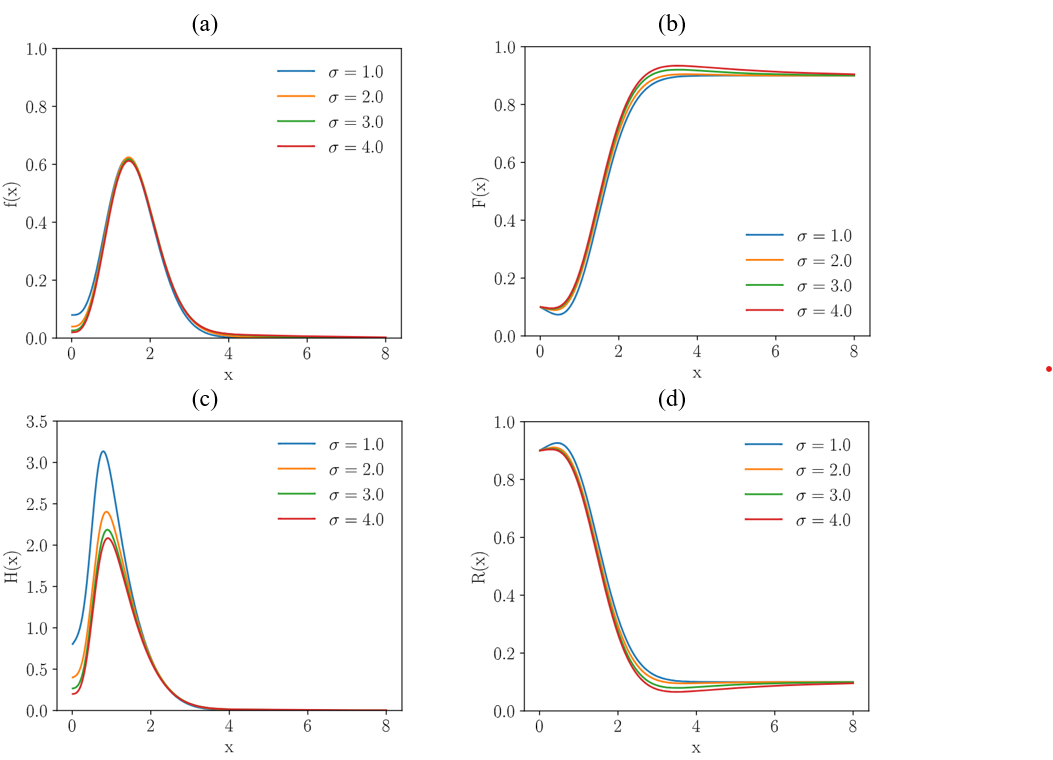}
 \caption{Comparison of EEPHND characteristics under varying $\sigma$ for fixed $\alpha = \theta = \lambda = \beta = 2$. Subplots: (a) PDF, (b) CDF, (c) Hazard, (d) Reliability, (e) Fitted PDF vs Data, (f) ECDF comparison.}
\label{fig:Fig1}
 \end{figure}

\begin{figure}[ht]
 \centering
  \includegraphics[width=1.0\textwidth]{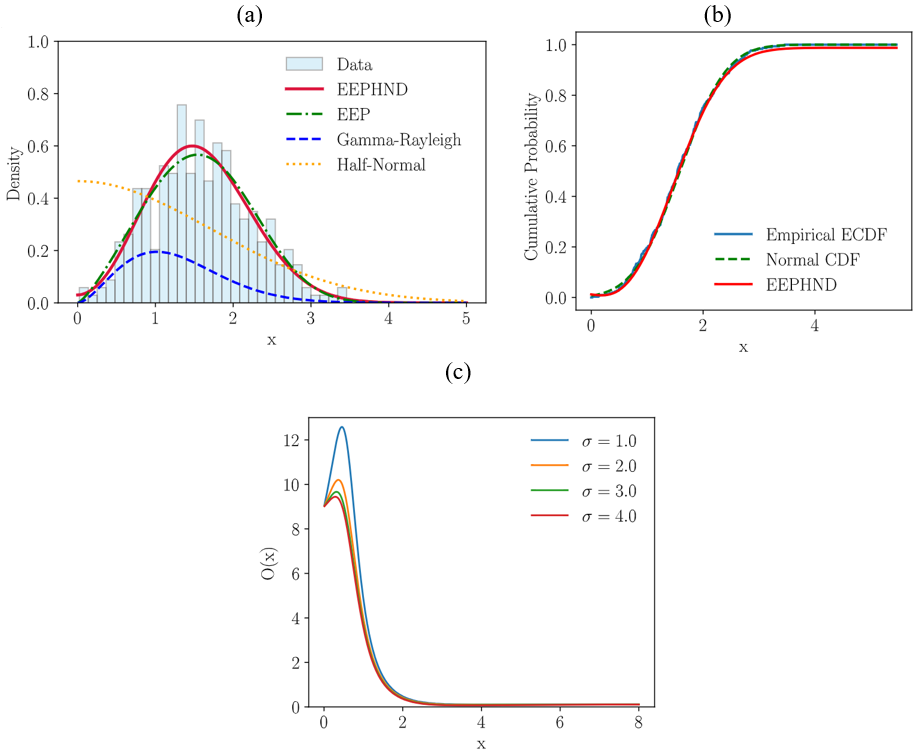}
 \caption{Comparison of different competing models (a), the CDF and EDCF plot (b) and the Odds function (c)}
\label{fig:Fig2}
 \end{figure}

Figure~\ref{fig:Fig1} shows the effect of varying $\sigma$ on the EEPHND with fixed $\alpha = \beta = \theta = \lambda = 2$. As $\sigma$ increases, the PDF broadens, the CDF smooths (Figure~\ref{fig:Fig2}), the hazard flattens, and reliability declines more gradually. panels (e) and (f) confirm that EEPHND fits simulated data well and better matches the ECDF than the normal model, highlighting its effectiveness for lifetime and reliability modeling.

The model comparison metrics (AIC, BIC, and CAIC) indicate that the EEPHND model has the lowest values, suggesting the best overall fit among the four models evaluated, closely followed by EEP, see Table~\ref{tab:model_comparison1}. In contrast, the Gamma-Rayleigh and HN models show significantly higher scores, indicating poorer fit. The estimated parameters for the EEPHND model include $\alpha = 1.1932$, $\beta = 1.8815$, $\theta = 2.4340$, $\lambda = 1.3058$, $\sigma = 0.3219$, and mixing probability $p_1 = 0.9878$, and$p_2 = 0.0122$ . The 95\% bootstrap confidence intervals for these parameters are reasonably tight, demonstrating parameter stability and suggesting that the EEPHND model provides a flexible and statistically reliable fit to the data.

\begin{table}[ht]
\caption{Model comparison metrics and estimated parameters with 95\% bootstrap confidence intervals.}
\label{tab:model_comparison1}
\centering
\begin{tabular}{llccc}
\toprule
\multicolumn{2}{l}{\textbf{Model Comparison (Lower is Better)}} & \textbf{AIC} & \textbf{BIC} & \textbf{CAIC} \\
\midrule
\multicolumn{2}{l}{EEPHND} & 599.6024 & 614.418 & 599.671 \\
\multicolumn{2}{l}{EEP}    & 600.024 & 622.247 & 600.167 \\
\multicolumn{2}{l}{GR}     & 1394.505 & 1401.913 & 1394.526 \\
\multicolumn{2}{l}{HN}     & 760.756 & 764.460 & 760.763 \\
\midrule
\multicolumn{5}{l}{\textbf{Estimated Parameters (MLE) and 95\% Bootstrap Confidence Intervals}} \\
\midrule
\textbf{Param} & \textbf{MLE} & \textbf{95\% CI Lower} & \textbf{95\% CI Upper} & \\
\midrule
$\alpha$  & 1.1932 & 0.5560 & 1.8465 & \\
$\beta$   & 1.8815 & 1.6253 & 2.1415 & \\
$\theta$  & 2.4340 & 1.9677 & 3.8680 & \\
$\lambda$ & 1.3058 & 0.6797 & 1.5823 & \\
$\sigma$  & 0.3219 & 0.0106 & 1.4331 & \\
$p_1$     & 0.9878 & 0.9261 & 0.9999 & \\
\bottomrule
\end{tabular}
\end{table}

\clearpage
\subsection{Validation with Real Data}

\subsubsection{Real Data}

The real-life clinical dataset used in this work is the publicly available lung cancer survival dataset from the lifelines Python package~\cite{davidson2019lifelines}. It can be accessed via the load$\_$lung() function, which returns a structured dataFrame containing time-to-event data, censoring indicators, and covariates such as age, sex, ECOG performance score, and weight loss. For reproducibility, all preprocessing steps, such as dropping missing values and rescaling the survival time to the range [0,1], were performed using standard Python pandas operations~\cite{mckinney2011pandas}. The dataset originates from the North Central Cancer Treatment Group~\cite{shaw2002prospective} and is frequently used for benchmarking survival models.

\begin{table}[h!]
\centering
\caption{Comparison of Survival Models: Cox PH, Kaplan-Meier, and EEPHND}
\resizebox{\textwidth}{!}{%
\begin{tabular}{|l|c|c|c|}
\hline
\textbf{Metric} & \textbf{Cox PH Model} & \textbf{Kaplan-Meier} & \textbf{EEPHND (Parametric)} \\
\hline
\multicolumn{4}{|c|}{\textbf{Concordance Index (CI)}} \\
\hline
CI Score & 0.6029 & 0.9982 & \textbf{0.9997} \\
\hline
\multicolumn{4}{|c|}{\textbf{Model Parameters}} \\
\hline
\textbf{Covariate: Age} & coef = 0.0170 & --- & --- \\
& $HR = 1.017$ & --- & --- \\
& 95\% CI = [-0.0010, 0.0351] & --- & --- \\
& $p$ = 0.0646 & --- & --- \\
\hline
\textbf{Covariate: Sex (M=1)} & coef = \textbf{-0.5132} & --- & --- \\
& $HR = \textbf{0.599}$ & --- & --- \\
& 95\% CI = \textbf{[-0.8414, -0.1850]} & --- & --- \\
& $p$ = \textbf{0.0022} & --- & --- \\
\hline
\multicolumn{4}{|c|}{\textbf{EEPHND Model Coefficients}} \\
\hline
$\alpha$ (shape) & --- & --- & 0.0001 (unstable) \\
$\beta$ (scale) & --- & --- & 0.02 \\
$\theta$ (exponent) & --- & --- & 0.01 \\
$\lambda$ (rate) & --- & --- & 13.79 \\
$\sigma$ (Half-Normal scale) & --- & --- & \textbf{0.46} \\
$p_1$ (EEP weight) & --- & --- & \textbf{0.01} \\
\hline
\multicolumn{4}{|c|}{\textbf{Survival Estimate at $t = 0.012$}} \\
\hline
$S(t)$ & 0.9791 & 0.9781 & $\approx 0.979$ \\
95\% CI & --- & [0.9481, 0.9908] & --- \\
\hline
\multicolumn{4}{|c|}{\textbf{Model Type and Flexibility}} \\
\hline
Type & Semi-parametric & Non-parametric & Fully parametric (mixture) \\
Baseline Hazard & Estimated & Stepwise constant & Closed-form from EEPHND \\
Handles Covariates & Yes & No & No (in current form) \\
\hline
\end{tabular}
}
\label{tab:survival_comparison}
\end{table}

\subsubsection{Model Comparison}

\begin{figure}[h]
 \centering
  \includegraphics[width=1.0\textwidth]{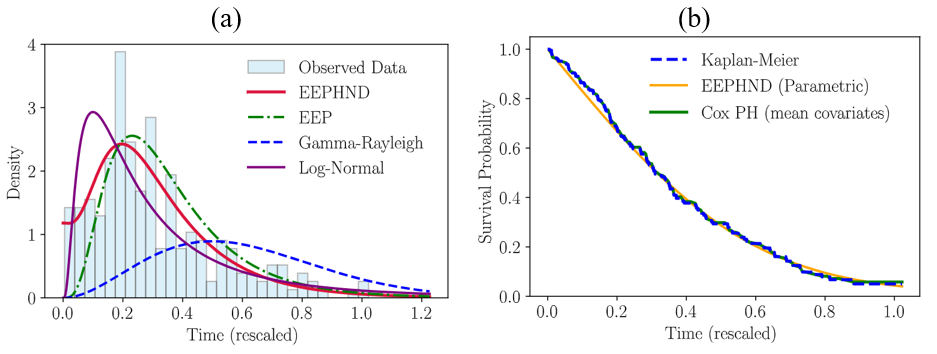}
 \caption{The density of different candidate models (a), and comparison of survival models (b) }
\label{fig:Survival}
 \end{figure}

Table~\ref{tab:survival_comparison} presents a comparative summary of three survival models, namely Cox Proportional Hazards (Cox PH), Kaplan-Meier, and the proposed EEPHND, applied to real clinical data. Based on the Concordance Index (CI), EEPHND (\textbf{CI = 0.9997}) and Kaplan-Meier (\textbf{CI = 0.9982}) clearly outperform the Cox PH model (CI = 0.6029), suggesting superior predictive accuracy. For covariates in the Cox PH model, the effect of age was marginally significant (p=0.0646), while sex (male) showed a statistically significant association with poorer survival (HR=0.599, p=0.0022). The EEPHND model, despite being fully parametric, does not incorporate covariates in its current formulation but produced highly flexible parameter estimates, notably a strong Half-Normal scale component ($\sigma$=0.46) and a very small EEP weight ($p_1$ = 0.01), see Figure~\ref{fig:Survival}. Survival estimates at t=0.012 were consistent across all models. Overall, EEPHND offers high accuracy and closed-form flexibility~\ref{tab:model_comparison2}, while the Cox model adds interpretability through covariate effects.

\begin{table}[h!]
\centering
\caption{Model Comparison using AIC, BIC, and CAIC criteria (lower is better)}
\begin{tabular}{|l|r|r|r|}
\hline
\textbf{Model} & \textbf{AIC} & \textbf{BIC} & \textbf{CAIC} \\
\hline
\textbf{EEPHND}                 & \textbf{-123.340} & \textbf{-102.764} & \textbf{-96.764} \\
Log-Normal             & -71.642  & -64.783  & -62.783 \\
EEP                    & 12.873   & 26.591   & 30.591 \\
Gamma-Rayleigh         & 497.744  & 504.603  & 506.603 \\
\hline
\end{tabular}
\label{tab:model_comparison2}
\end{table}

The EEPHND model captures both early- and late-risk populations through a mixture of EEPD and HN components, governed by a flexible mixture parameter $p_1$ with $p_2 = 1-p_1$. Compared to Cox PH and Kaplan-Meier models, EEPHND provides an explicit parametric form for the baseline survival, allowing direct interpretation of shape, scale, and risk mixture in survival dynamics. Notably, the heavy tail behavior (modulated by $\alpha$, $\theta$, and $\lambda$) makes it well-suited for datasets with heterogeneous risk profiles, unlike Cox PH models which rely heavily on covariates~\ref{tab:survival_comparison} for capturing such nuances.

Overall, the Cox model identifies sex as a significant predictor, with males showing a lower hazard rate than females (HR = 0.599, p = 0.0022), highlighting the value of covariate inclusion in detecting clinical risk differences. While the EEPHND model currently lacks covariate, it outperforms both Cox and Kaplan-Meier in predictive accuracy (CI = 0.9997) and closely matches their early survival estimates. Its fully parametric, closed-form structure enables continuous survival modeling, making it well-suited for clinical tasks like prognosis simulation and population-level planning, particularly when individual covariate data are limited or not available. Together, these models offer complementary strengths in risk interpretation and survival forecasting.

\section{conclusion}

This work propose EEPHND model as a flexible, interpretable model for complex survival and reliability data. By combining the EEP and HN distributions, EEPHND captures both early-time symmetry and long-tail behavior often missed by classical or generalized models. Theoretical development was supported by empirical validation on a real-world lung cancer dataset, where EEPHND outperformed standard models in both goodness-of-fit and predictive accuracy, achieving a Concordance Index of 0.9997. These results affirm its utility as a robust tool for lifetime data modeling and its broad relevance to biomedical research, and other fields requiring nuanced risk assessment.



\section*{Author Contribution}

Oriyomi Ahmad Hassan: Writing – review $\&$ editing, Validation, Software, Methodology, Investigation, Formal analysis, Data curation, Conceptualization. Aisha Tunrayo Maradesa: Writing – review $\&$ editing, Writing – original draft, Software, Methodology. Toyosi Abdulazeez Alabi: Writing – review $\&$ editing, Visualization, Validation, Software, Methodology, Investigation.  Oyejide Surajudeen Salam: Writing – review $\&$ editing, Validation, Investigation. Ajani Busari: Writing – review $\&$ editing, Investigation. Akinwale Victor Famotire: Writing – review $\&$ editing, Methodology, Visualization, Investigation. Habeeb Abiodun Afolabi: Writing – review $\&$ editing, Investigation. Solomon Adeleke: Writing – review $\&$ editing, Visualization. Abayomi Ayodele Akomolafe: Supervision, Software, Project administration, Methodology, Investigation, Conceptualization.

\clearpage
\bibliography{main.bbl}

\end{document}